\documentclass[aps,preprint,groupedaddress]{revtex4-2}
\usepackage{amsmath}
\usepackage{graphicx}
\usepackage{graphics}
\usepackage{natbib}
\usepackage{float}
\usepackage{color,xcolor}
\usepackage{cancel}  
\usepackage{amssymb}
\usepackage{titlesec}

\begin{document}
\title{Ultrasensitive refractive index sensor with rotatory biased weak measurement}
\author{Chongqi Zhou}
\affiliation{Department of Physics, Tsinghua University, Beijing 100084, China}
\affiliation{Institute of Optical Imaging  and Sensing, Shenzhen Key Laboratory for Minimal Invasive Medical Technologies, Shenzhen International Graduate School, Tsinghua University, Shenzhen 518055, China}
\author{Yang Xu}
\affiliation{Institute of Optical Imaging  and Sensing, Shenzhen Key Laboratory for Minimal Invasive Medical Technologies, Shenzhen International Graduate School, Tsinghua University, Shenzhen 518055, China}
\author{Xiaonan Zhang}
\affiliation{Institute of Optical Imaging  and Sensing, Shenzhen Key Laboratory for Minimal Invasive Medical Technologies, Shenzhen International Graduate School, Tsinghua University, Shenzhen 518055, China}
\author{Zhangyan Li}
\altaffiliation{}
\affiliation{Department of Physics, Tsinghua University, Beijing 100084, China}
\affiliation{Institute of Optical Imaging  and Sensing, Shenzhen Key Laboratory for Minimal Invasive Medical Technologies, Shenzhen International Graduate School, Tsinghua University, Shenzhen 518055, China}
\author{Tian Guan}
\affiliation{Institute of Optical Imaging  and Sensing, Shenzhen Key Laboratory for Minimal Invasive Medical Technologies, Shenzhen International Graduate School, Tsinghua University, Shenzhen 518055, China}
\author{Yonghong He}
\affiliation{Institute of Optical Imaging  and Sensing, Shenzhen Key Laboratory for Minimal Invasive Medical Technologies, Shenzhen International Graduate School, Tsinghua University, Shenzhen 518055, China}
\author{Yanhong Ji}
\email{jiyh@scnu.edu.cn}
\affiliation{School of Physics and Telecommunication Engineering, South China Normal University, Guangzhou 510006, China}

\begin{abstract}
A modified weak measurement scheme, rotatory biased weak measurement, is proposed to significantly improve the sensitivity and resolution of the refractive index sensor on a total reflection structure. This method introduces an additional phase in the post-selected procedure and generates an extinction point in the spectrum distribution. The biased post-selection makes smaller coupling strength available, which leads to an enhancement of phase sensitivity and refractive index sensitivity. In rotatory biased weak measurement, we achieve an enhanced refractive index sensitivity of 13605 nm/RIU compared to 1644 nm/RIU in standard weak measurement. The performance of sensors with different sensitivity is analyzed, and we find the optimal refractive index resolution of sensors increases with sensitivity. In this work, we demonstrate an optimal refractive index resolution of $4\times10^{-7}$ RIU on a total reflection structure. The rabbit anti-mouse IgG and mouse IgG binding reaction experiments demonstrate that our system has a high response to the concentration of IgG in a wide range and the limit of detection is 15 ng/mL. The improvements in this work are helpful to the optimizations of other optical sensors with weak measurement.

\end{abstract}
\maketitle
\newpage
\section{Introduction}
Optical sensors are widely used in chemical, biomedical and food fields as refractive index (RI) sensors, such as total internal reflection, surface plasmon resonance (SPR), fiber interferometers, and metal-cladding waveguide sensors \cite{goodling2019colouration,gao2015reconfigurable,ma2019situ,xu2019optical,quan2015ultra,wang2008oscillating,armani2007label}. In the past decades, researchers have proposed various structures, such as long-range SPR \cite{slavik2007ultrahigh} and waveguide coupling sensors \cite{mishra2016surface}, to improve the RI sensitivity and resolution of sensors.  For example, long-range SPR improves the sensitivity of the system by enhancing the electromagnetic field at the interface; total internal reflection sensors use high-temperature reduced graphene oxide to amplify the light intensity response to RI \cite{xing2014ultrasensitive}. However, the sensitivity of these sensors is determined by their structures and materials. And it is difficult for us to further enhance their performance on the existing sensing structures and materials. Therefore, how to achieve enhanced sensitivity and resolution of sensors in a convenient way is an interesting problem worth investigating. 

Standard weak measurement and weak value amplification, proposed by Aharonov \cite{aharonov1988result}, is an innovative strategy for measuring small physical quantities. Unlike conventional measurement, standard weak measurement amplifies the contrast signal by introducing post-selection. Over the past decades, standard weak measurement has been widely used in high precision measurement \cite{jozsa2007complex,li2017adaptive,kedem2012using,brunner2010measuring,pang2016protecting,hosten2008observation} and provided much higher sensitivity for the measurement of small physical effects than previous techniques. In 2008, the spin Hall effect is first observed with standard weak measurement, where the beam offset is amplified by a factor of 100, achieving a resolution of 1 $\mathring{\mathrm{A}}$ \cite{hosten2008observation}. Subsequently, standard weak measurement is experimentally applied to measurements of other physical parameters, such as phase \cite{xu2013phase}, temperature \cite{li2018high}, and optical nonlinearity \cite{li2019ultra}.  In these experiments, standard weak measurement has achieved higher sensitivity and resolution with post-selection. Similarly, based on standard weak measurement, we have proposed a RI sensor with a RI resolution of $3.6\times 10^{-6}$ RIU \cite{zhang2016optical}, which is superior to the total internal reflection sensor \cite{azzam2004phase}. In an ideal case, standard weak measurement can achieve high sensitivity and resolution with stronger post-selection. However, in practice, the post-selection in standard weak measurement cannot be arbitrarily strong, and the performance of the system can be only enhanced to a limited extent \cite{xu2020approaching,xu2018optimization}. By contrast, biased weak measurement can attain stronger post-selection by introducing an additional phase in the post-selected procedure, which makes the system more sensitive to RI.  This allows us to further improve the performance of the sensor on the existing sensor.

In this work, we propose an ultrasensitive RI sensor based on rotatory biased weak measurement, where the coupling interaction is performed by the optical rotatory effect. It allows us to introduce an additional phase in the post-selection process, and further enhance the performance of the sensor. Here, we implement RI sensors with different sensitivities in the same structure and measure their optimal RI resolution. It is proved that the optimal RI resolution increases with the RI sensitivity. Using the optimal system, we have achieved an enhanced RI sensitivity of 13605 nm/RIU and RI resolution of $4\times 10^{-7}$ RIU. Finally, the binding reaction of anti-IgG and IgG is performed to demonstrate the detection capability of our biosensor. 

\section{Material and methods}
\subsection{Materials}
Equilateral prisms (Chinese ZF6 glass, $27\times 27\times 27$mm) are ordered from Fuzhou Alpha Optics Co., Ltd. (Fuzhou, China). Phosphate buffered saline (PBS) powder is purchased from Shanghai RuiChu Biotech Co., Ltd. (Shanghai, China). Dopamine hydrochloride and Tris (hydroxymethyl) aminomethane are bought from Aladdin Industrial Corporation (Shanghai, China). No protein blocking solution is bought from Sangon Biotech Co., Ltd. (Shanghai, China). Mouse immunoglobulin (IgG) and rabbit anti-mouse IgG are obtained from Bioss (Beijing, China). NaCl is ordered from Aladdin Industrial Corporation (Shanghai, China). All other chemical reagents are purchased from Shenzhen Tianxiang Huabo Co., Ltd. (Shenzhen, China).

\begin{figure}[ht!]
\centering
\includegraphics[width=\linewidth, trim=0 0 0 0, clip]{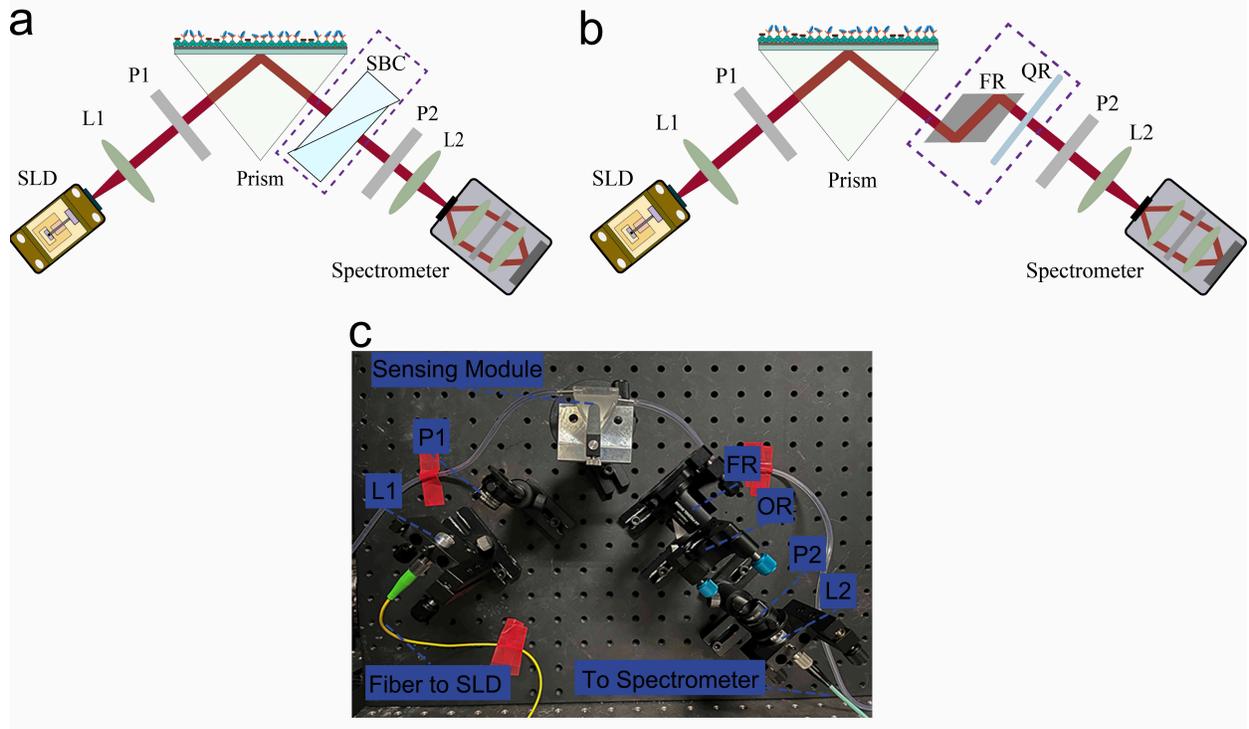}
\caption{Schematic illustration of weak measurement schemes and experimental setup. (a) Schematic of RI sensor with standard weak measurement. SLD, Superluminescent Diodes; P1 and P2, linear polarizers; L1 and L2, lens; SBC, Soleil-Babinet compensator. (b) Schematic of RI sensor with rotatory biased weak measurement. FR, Fresnel rhomb retarder. QR, quartz rotator. (c) Photograph of a RI sensing setup with biased weak measurement. In the sensing module, samples are pumped through the flow cell by a Syringe pump (pump 11, Harvard Apparatus Co., Ltd., USA). }
\label{fig1}
\end{figure}

\subsection{Principles of biased weak measurement and standard weak measurement}

\begin{figure}[ht!]
\centering
\includegraphics[width=0.5\linewidth, trim=0 0 0 0, clip]{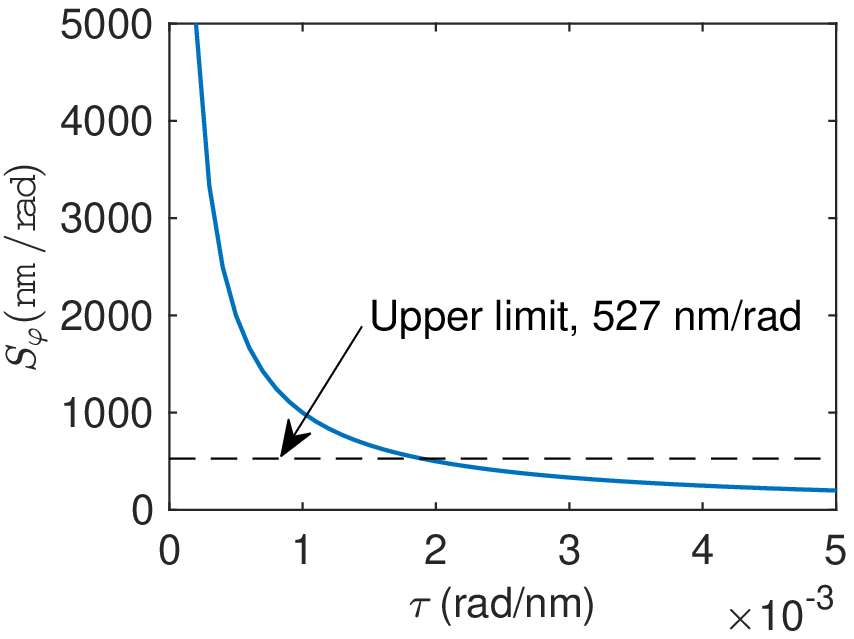}
\caption{The relationship between the phase sensitivity $S_{\varphi}$ and the coupling strength $\tau$. In the weak measurement scheme, the phase sensitivity of the system  $S_{\varphi}$ is proportional to $1/\tau$. In our system, $\lambda_0=833$ nm. The upper limit of phase sensitivity that can be achieved by standard weak measurement is 527 nm/rad.}
\label{fig2}
\end{figure}

When the light is totally reflected by the prism at the incident angle of $\theta$, a phase difference $\varphi$ is introduced between the $p$ and $s$ polarizations. According to Fresnel's formula, we have
\begin{equation}
    \varphi=2\mathrm{tan}^{-1}\frac{\sqrt{n_{1}^2\mathrm{sin}^2\theta-n_{2}^{2}}}{n_{1} \mathrm{sin}\theta \mathrm{tan}\theta},
    \label{eq8}
\end{equation}
where $n_{1}=1.75$ and $n_{2}\approx1.33$ are refractive indices of the prism and analytes, respectively. We consider a weak measurement scheme to measure the phase difference $\varphi$, which varies with the RI of the analytes. Normally, weak measurement consists of three parts: pre-selection, weak coupling, and post-selection. For biased weak measurement, as shown in Fig. \ref{fig1}b, the preselection is completed by polarizer P1, prism, and Fresnel retarder. Therefore, the preselected polarization state can be written as 
\begin{equation}
|\psi_{\mathrm{pre}}\rangle=\frac{1}{\sqrt{2}}(e^{i(\frac{\varphi}{2}-\frac{\pi}{4})}|\circlearrowright\rangle+e^{-i(\frac{\varphi}{2}-\frac{\pi}{4})}|\circlearrowleft\rangle),
\end{equation}
in which $|\circlearrowright\rangle=(|p\rangle+i|s\rangle)/\sqrt{2}$ and $|\circlearrowleft\rangle=(|p\rangle-i|s\rangle)/\sqrt{2}$ represent right-handed and left-handed circular polarizations, respectively.

In rotatory biased weak measurement system, the coupling process is completed by quartz cut perpendicular to the optic axis, which causes optical rotation or circular birefringence. The rotatory power varies with the wavelength, which is known as the rotatory dispersion effect \cite{moffitt1961structure,zhou2020measuring}. The interaction between the system (polarization) and the meter (wavelength spectrum) can be described by the unitary transformation
\begin{equation}
U=e^{-\tau\delta(t-t_0)\hat{\mathbf{A}}\hat{\lambda}},
\end{equation}
where $\hat{\mathbf{A}}=|\circlearrowright\rangle\langle \circlearrowright|-|\circlearrowleft\rangle\langle \circlearrowleft|$ and $\hat{\lambda}$ is the wavelength operator of the photon. The meter $\int f(\lambda)|\lambda\rangle d\lambda$ is assumed to have a Gaussian profile with center wavelength $\lambda_0$ and variance $\sigma_0$
\begin{equation}
f(\lambda)=(\sqrt{\pi}\sigma_0)^{-1/2}e^{-\frac{(\lambda-\lambda_{0})^2}{2\sigma_{0}^2}}.
\end{equation}
So the joint state after coupling can be written as 
\begin{widetext}
\begin{align}
|\Psi\rangle&=\frac{1}{\sqrt{2}}\int d\lambda(e^{i(\frac{\varphi}{2}-\frac{\pi}{4}+\tau\lambda)}|\circlearrowright\rangle+e^{-i(\frac{\varphi}{2}-\frac{\pi}{4}+\tau\lambda)}|\circlearrowleft\rangle)|f(\lambda)|\lambda\rangle.
\label{eqn4}
\end{align}
\end{widetext}

The joint state is projected on the post-selected state by the polarizer P2. It can be described as
\begin{equation}
|\psi_{\mathrm{post}}\rangle=\frac{1}{\sqrt{2}}(e^{i(\frac{\pi}{4}+\varepsilon)}|\circlearrowright\rangle+e^{-i(\frac{\pi}{4}+\varepsilon)}|\circlearrowleft\rangle).
\end{equation}
As the system is projected on the postselected state, it leads to a nonnormalized distribution of wavelength to be
\begin{equation}
    D(\lambda)=\mathrm{sin}^2(\tau\lambda+\frac{\varphi}{2}-\varepsilon)|f(\lambda)|^2.
\end{equation}
And the shift of central wavelength $\delta\lambda$ in the reverse weak value regime $|\tau\lambda_0-\varepsilon+{\varphi}/{2}|\ll\tau\sigma_0$ \cite{dressel2013strengthening} can be written as \begin{align}
    \delta\lambda &=\frac{\int D(\lambda)\lambda d\lambda}{\int D(\lambda) d\lambda}-\lambda_{0}\notag\\
&=\frac{2\tau\sigma_0^2(2\tau\lambda_0+\varphi-2\varepsilon)}{2\tau^2\sigma_0^2+(2\tau\lambda_0+\varphi-2\varepsilon)^2}\notag\\
   &\approx2\lambda_0+\frac{\varphi-2\varepsilon}{\tau},\ |\tau\lambda_0-\varepsilon+\frac{\varphi}{2}|\ll\tau\sigma_0.
\label{eq15}
\end{align}

For standard weak measurement, as shown in Fig. \ref{fig1}a, the initial state is generally set as $(|p\rangle+|s\rangle)/\sqrt{2}$. The interaction between the system and the meter is introduced by wave plate, where the Hamiltonian can be written as $H=\tau\delta(t-t_0)\hat{\lambda}\hat{B}$, in which $\hat{B}=|p\rangle\langle p|-|s\rangle\langle s|$. A normal post-selection into $|\psi\rangle=((|p\rangle-|s\rangle)/\sqrt{2})$ is completed by a polarizer, and the distribution of wavelength after post-selection is given as 
\begin{equation}
    D'(\lambda)=\mathrm{cos}^{2}(\tau\lambda+\frac{\varphi}{2})f(\lambda).
\end{equation}

Therefore, the value of $\varphi$ can be estimated through the shift of the central wavelength $\delta\lambda'$
\begin{equation}
    \delta\lambda'=\frac{\int D'(\lambda)\lambda d\lambda}{\int D'(\lambda) d\lambda}-\lambda_{0}\approx2\lambda_0+\frac{\varphi}{\tau},
\label{eq16}
\end{equation}
in which the signal is amplified by a factor of $\tau$ in the reverse weak-value regime $|\tau\lambda_0+\varphi/2|\ll\tau\lambda_0$. 

It has been proved that the weak measurement system is extremely sensitive to $\varphi$ at the reverse weak-value regime \cite{dressel2013strengthening,zhou2020measuring}. For standard weak measurement, it imposes a rigorous restriction on the value of $\tau$ as it satisfies $\tau\simeq(m+1/2)\pi/\lambda_0$. As shown in Eq. (\ref{eq15}) and (\ref{eq16}), the shift of the central wavelength is amplified by the factor of $1/\tau$. And the highest phase sensitivity of $2\pi/\lambda_0\approx527$ nm/rad can be obtained for standard weak measurement, which prevents further enhancement of sensitivity with weak measurement. However, compared to standard weak measurement, biased weak measurement introduces an additional phase in the post-selection and gives a further amplified signal of central wavelength shift $\delta\lambda$. In biased weak measurement, it requires $|\tau\lambda_0-\varepsilon+\frac{\varphi}{2}|\approx0$. This limitation can be accomplished for arbitrary $\tau$ by adjusting the value of $\varepsilon$. As a result, biased weak measurement can acquire a higher phase sensitivity than standard weak measurement when $\tau<\pi/\lambda_0$.

\subsection{System setup}

The experimental setup is depicted in Fig. \ref{fig1}c. The broadband light source used here is a superluminescent laser diode (SLD, IPSDD0804, 15 mW, Inphenix). The beam transmitted through the optical fiber is collimated by the achromatic lens L1 and preselected by polarizer P1(Thorlabs Inc., 180 LPVIS050-MP, extinction ratio of 100 000:1), whose transmission axis is $45 ^{\circ}$ from the vertical direction. Then the beam is totally reflected on the inner surface of the ZF6 prism, resulting in a  RI-independent phase difference $\varphi$ between $s$ and $p$ polarizations. Samples are pumped through the flow cell by a Syringe pump (pump 11, Harvard Apparatus Co., Ltd., USA). To transform the phase difference between $s$ and $p$ polarizations into optical rotation, the fast axis of the Fresnel rhomb retarder FR (Thorlabs Inc., FR600QM) is set at $45 ^{\circ}$ from the vertical direction. The rotation angle is proportional to the phase difference $\varphi$.  After that, the beam passes through quartz crystal along the optical axis, which finally is postselected by the second polarizer P2. The postselected light is collected by a spectrograph (OCEANVIEW, HR4000, Shanghai, China) for spectral analysis.  

\section{Results and discussions}

\subsection{Improving phase sensitivity with rotatory biased weak measurement}

\begin{figure}[ht!]
\centering
\includegraphics[width=\linewidth, trim=0 0 0 0, clip]{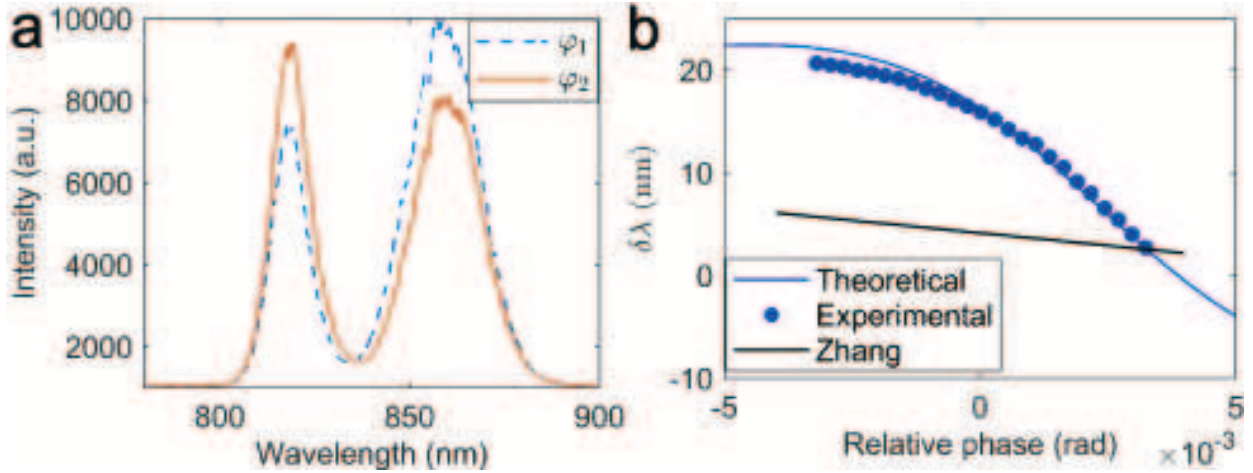}
\caption{Experimental results of biased weak measurement (a) Spectral signal changes with phase $\varepsilon$. In the inverse weak-value regime, the change of phase results in the shift of the extinction point. (b) Experimental and theoretical results of central wavelength shift versus phase. Compared with standard weak measurement in Zhang's scheme, the phase sensitivity of biased weak measurement is improved to 4478 nm/rad, which is one order of magnitude higher than that of standard weak measurement.}
\label{fig3}
\end{figure}

In order to compare the phase sensitivity of biased weak measurement and standard weak measurement system, we measure the shift of central wavelength with $\varepsilon$ for biased weak measurement and standard weak measurement. In the biased weak measurement, we use a 0.3 mm-thickness quartz crystal as the coupling process with a coupling strength of $2\times10^{-4}$ rad/nm. The incident angle of total reflection is set to $52^{\circ}$. In this system, the relationship between the shift of central wavelength and phase can be achieved by rotating the rear polarizer. As shown in Fig. \ref{fig3}a, the extinction point in the spectrum moves with the phase $\varepsilon$. We rotate the polarizer with a step of $2.7\times 10^{-4}$ rad and measure central wavelength shift, as shown in Fig. \ref{fig3}b. In the linear region, the phase sensitivity of the system is 4478 nm/rad. Meanwhile, the spectral probability distribution of the light source is measured to be a superposition of two Gaussian distributions (not normalized)
$ |f(\lambda)|^2=\mathrm{exp}[{-{(\lambda-\lambda_{1})^2}/{\sigma_{1}^2}}]+ 1.035\times \mathrm{exp}[{-{(\lambda-\lambda_{2})^2}/{\sigma_{2}^2}}]$, in which  $\lambda_{1}=821.1$ nm, $\sigma_{1}=7.55$ nm, $\lambda_{2}=845.8$ nm and $\sigma_{2}=19.58$ nm. According to Eq. (\ref{eq15}), the theoretical relationship between central wavelength shift $\delta\lambda$ and phase $\varphi$ is calculated. Fig. \ref{fig3}b demonstrates a great correspondence between theoretical and experimental results. For the standard weak measurement in Zhang's scheme, the phase sensitivity is calibrated as 527 nm/rad. Therefore, the biased weak measurement achieves a 9 times higher phase sensitivity than that of the standard weak measurement.

\subsection{Optimization of RI sensitivity}

\begin{figure}[ht!]
\centering
\includegraphics[width=\linewidth, trim=0 0 0 0, clip]{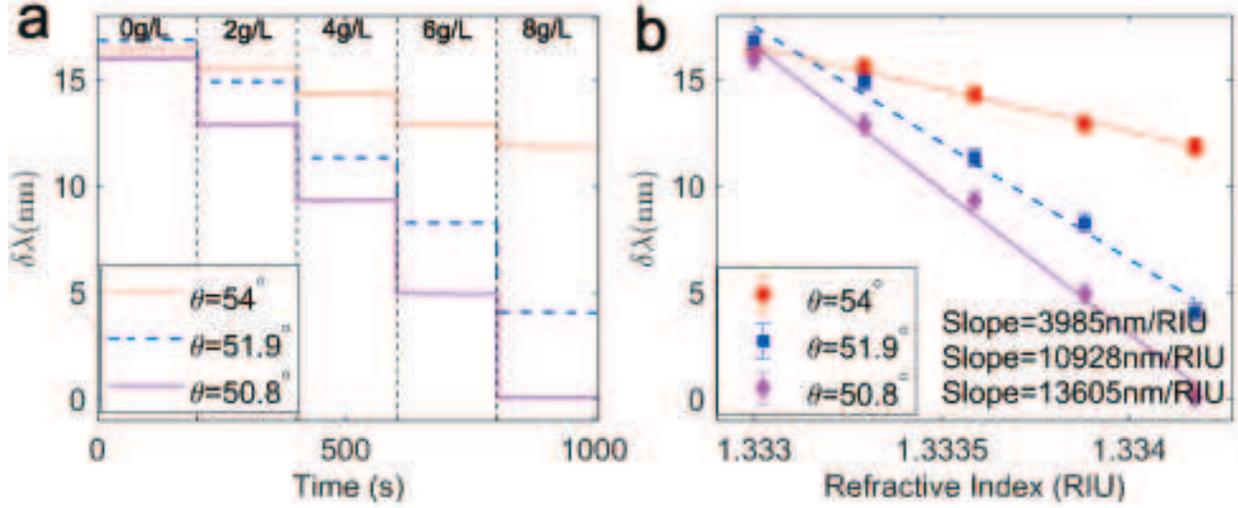}
\caption{Optimization of the RI sensitivity of the biased weak measurement system. (a) Central wavelength shift corresponds to different concentrations of NaCl solutions at three incidence angles. (b) Linear fit of the central wavelength shift and RI at three incident angles. The highest RI sensitivity is 13605 nm/RIU at an incidence angle of $50.8^{\circ}$.}
\label{fig4}
\end{figure}

As shown in Eq. (\ref{eq8}), the relationship between phase and RI is influenced by the incident angle. As the phase sensitivity of the system is determined, the RI sensitivity of the system can be further optimized by changing the incident angle. Here, RI sensitivities of the system at three incident angles of $50.8^{\circ}$, $51.9^{\circ}$, $54^{\circ}$ are obtained. The sodium chloride solutions with different concentrations are employed to play the role of the variation of the RI. The relationship between the RI $n$ and the concentration of the sodium chloride solution ($C$,g/L) is calibrated by an Abbe refractometer as $n=1.3305+1.471\times10^{-4}C$. Sodium chloride solutions with concentrations of 0, 2, 4, 6, and 8 g/L are pumped into the flow cell to calibrate the sensor system and 200 spectra are collected for each solution. Fig. \ref{fig4}a shows the step change of central wavelength shift with different concentrations at three different angles. From the average value of each step, we can obtain the relationship between the central wavelength shift and the RI of the system. As shown in Fig. \ref{fig4}b, the central wavelength shift at different angles varies linearly with the RI, whose slope is the RI sensitivity of the system. As the incident angle approaches the total reflection angle ($49.6^{\circ}$), the RI sensitivity of the system gradually increases. In this system, it obtains the highest RI sensitivity of 13605 nm/RIU when $\theta=50.8^{\circ}$.
\subsection{Optimization of RI resolution of two schemes}
\begin{figure}[ht!]
\centering
\includegraphics[width=0.8\linewidth, trim=0 0 0 0, clip]{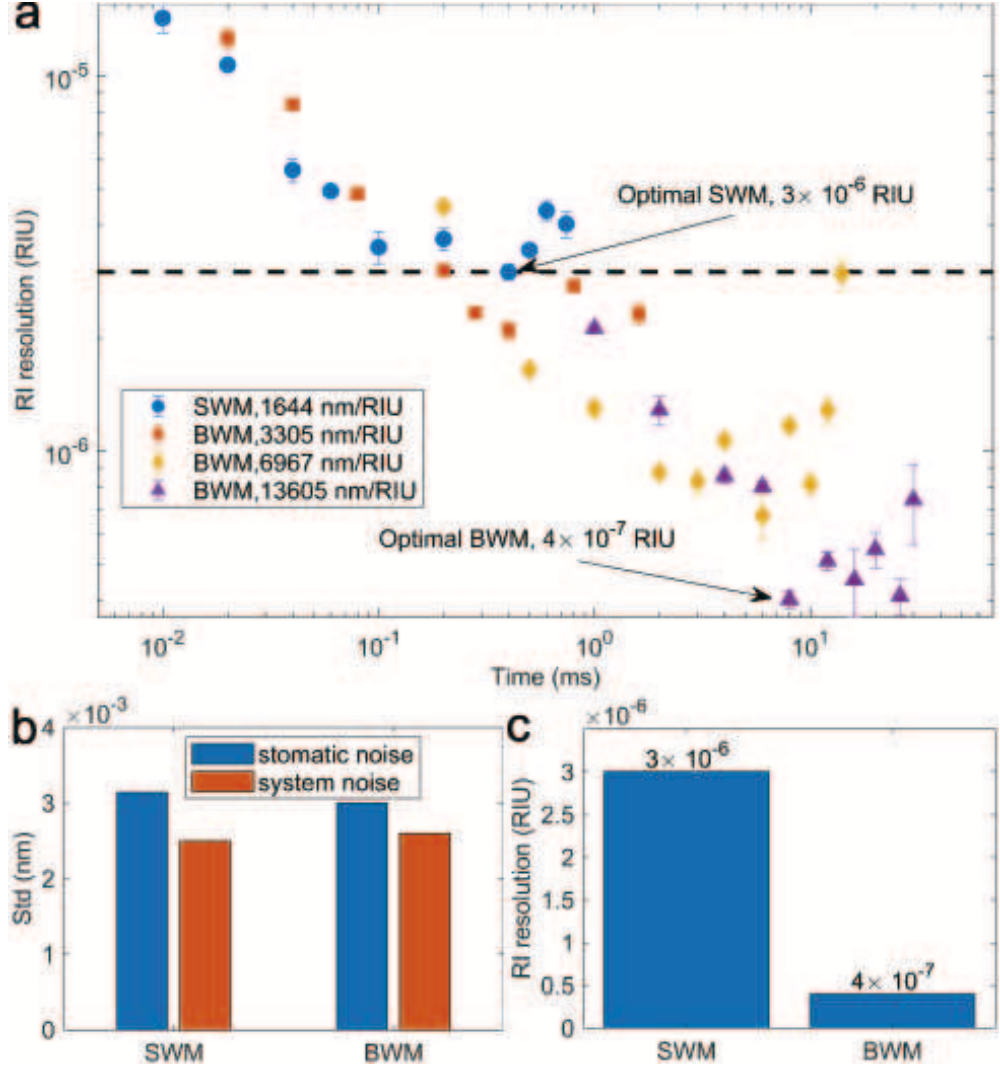}
\caption{Optimization of RI resolution for weak measurement systems. (a) The RI resolution versus response time for the weak measurement schemes with different RI sensitivities. Among them, the RI sensitivity of the standard weak measurement is 1644 nm/RIU, and the sensitivities of the biased weak measurement are 3305 nm/RIU, 6967 nm/RIU, and 13605 nm/RIU, respectively. The optimal RI resolution of the system decreases with RI sensitivity. (b) Noise analysis of the optimal standard weak measurement and biased weak measurement. As the response time increases, the stomatic noise is close to the system noise. It is difficult to reduce the RI resolution of the system by time averaging. (c) RI resolution of the optimal standard weak measurement and biased weak measurement. The biased weak measurement attain higher RI resolution with a RI sensitivity of 13605 nm/RIU.  }
\label{fig5}
\end{figure}
The RI resolution $r_{RI}$ of the system can be expressed as
\begin{equation}
r_{RI}=\frac{\sigma}{s_{RI}}.
\end{equation}
Detection noise $\sigma$ is classified into two categories: stomatic noise $\sigma_s$, such as light intensity jitter of the light source and electronic noise of the detector, and system noise  $\sigma_c$ like environmental perturbation and technical noise of the system. In a single acquisition, the stomatic noise occupies the major part of the noise and can be optimized by time averaging at the expense of temporal resolution. However, time-averaging doesn't diminish the system noise, and it becomes the dominant noise limiting the system. Therefore, the detection noise of the system versus averaging times $N$ can be written as
\begin{equation}
r^N_{RI}=\frac{\sqrt{\sigma^2_s/N+\sigma^2_c}}{s_{RI}}.
\label{eq17}
\end{equation}
Fig. \ref{fig5}a depicts the RI resolution versus response time for weak measurement systems with four coupling strengths.  and we measure the RI resolution of the system with response time by time averaging. Among them, the RI sensitivity of the standard weak measurement is 1644 nm/RIU and the sensitivities of the biased weak measurement are 3305 nm/RIU, 6967 nm/RIU, and 13605 nm/RIU, respectively. As illustrated in Fig. \ref{fig5}a, the RI resolution of standard weak measurement decreases sharply with the response time from 10 us to 100 us. As the stomatic noise decreases with averaging time, the RI resolution of the system no longer improves with the response time from 100 us to 800 us. And the optimal RI resolution achieved by standard weak measurements is $3\times10^{-6}$ RIU. However, biased weak measurement can further improve the RI resolution of the system by increasing the RI sensitivity, as shown in Eq. (\ref{eq17}). The time response is longer due to the stronger post-selection of the biased weak measurement. 

To calibrate the stomatic noise and system noise in the system, we consider that the readout of each pixel in the 3648-element CCD (Toshiba, TCD1304AP) can be written as
\begin{equation}
 k_j(t)= \bar{n}_j+c_j(t)+s_j(t),
\end{equation}
where $\bar{n}_j$ is the mean of intensity, $c_j(t)$ is the system noise and $s_j(t)$ is the stochastic component of the noise. We consider three effects in the model of stomatic noise. The first effect is the number of photoelectrons at the $j$th pixel on the CCD (3648 element), which follows Poisson distrubution $p(\bar{n}_j)$.  The second effect is the dark noise of CCD, which satisfies the normal distribution $N(k_d,\sigma_d)$, where $k_d=1040$ and $\sigma=4.5$. The third effect in the classical noise, which follows the noise distribution $N(0,\sigma^l_i)$, in which $\mathrm{ln}\sigma^l_i=0.46\bar{n}_j-1.58$.
The system noise is assumed to be independent of the stomatic noise and has a mean value of 0. According to Eq. (\ref{eq15}), the stomatic noise of the system can be written as 
\begin{align}
\sigma_s^N=\sqrt{\sum_{i=1}^{3648}(\frac{\sum_{j=1}^{3648} \bar{n}_j(\lambda_i-\lambda_j)}{(\sum_{j=1}^{3648} \bar{n}_j)^2})^2(\bar n_i^2+20.25+(\sigma^l_i)^2)}.
\label{eq14}
\end{align}

Fig. \ref{fig5}b shows the stomatic noise of the optimal standard weak measurement and the biased weak measurement according to Eq. (\ref{eq14}). The system noise can be calculated by subtracting the stomatic noise from the detection noise. In the two optimal weak measurement schemes, the value of the system noise is close to that of the stomatic noise. At this point, the performance of the system is mainly determined by the RI sensitivity. Therefore, the biased weak measurement can achieve higher RI resolution compared to the standard weak measurement. And an optimal RI resolution of $4\times10^{-7}$ RIU (response time: 7 ms) and  $1.7\times10^{-7}$ RIU (response time: 5 s) is obtained for the biased weak measurement system with the RI sensitivity of 16305 nm/RIU.

\begin{figure}[ht!]
\centering
\includegraphics[width=0.75\linewidth, trim=0 0 0 0, clip]{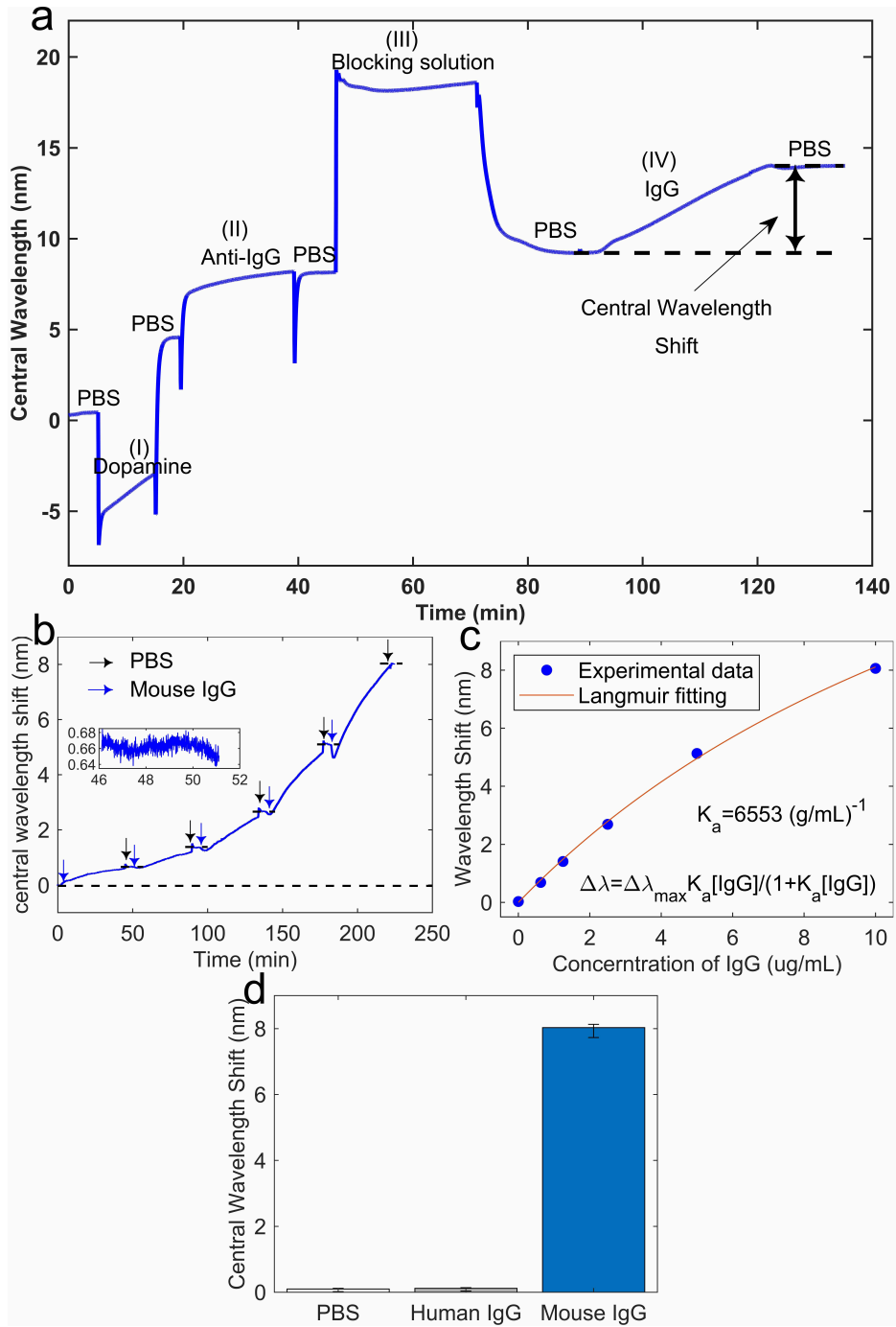}
\caption{Bio-experiments of IgG and anti-IgG binding reactions. (a) Schematic diagram of the binding reaction in the whole process of surface functionalization (I), rabbit anti-mouse IgG immobilization (II), protein-free blocking (III), and mouse IgG binding reaction (IV). (b) Binding kinetic curves of mouse IgG with different concentrations. PBS is circulated for 10 min after mouse IgG with different concentrations to obtain a baseline. (c) Langmuir fitting curve of the antibody-antigen binding interaction. (d) The specific binding ability of the sensor.}
\label{fig6}
\end{figure}

\subsection{Bio-experiment}

To demonstrate the sensing capability of the sensor, the binding reaction of rabbit anti-mouse IgG and mouse IgG is performed at the optimal incident angle of $50.8^{\circ}$. The solution flows over the surface of the prism through a flow cell, which is 2 cm long and 2 mm wide by 1 mm deep. The process of functionalization, antibody immobilization, and binding reactions on the prism surface is demonstrated in Fig. \ref{fig6}a. After 10 min circulation of PBS solution to obtain a stable baseline signal, 1 mg/mL of dopamine solution in 10 mM Tris-HCl buffer is injected for 10 min to form a thin adherent layer on the prism by self-polymerization reaction, which can be used for the multifunctional binding platform. After washing by the injection of PBS solution for 10 min, 100ug/mL rabbit anti-mouse IgG in PBS buffer is injected for 30 min to immobilize a biorecognition layer. After washing by the injection of PBS solution for 10 min, a 500 ug/mL protein-free blocking solution is injected for 20 min to block the uncombined polydopamine sites. After washing by the injection of PBS solution for 10 min, IgG solution in PBS buffer is injected for 50 min, followed by 10 min circulation of PBS solution. The central wavelength shift caused by the specific recognition of rabbit anti-mouse IgG and mouse IgG is demonstrated in Fig. \ref{fig6}a. 

After the functionalization and antibody immobilization on the prism surface, the sensor system is employed to detect mouse IgG. As illustrated in Fig. \ref{fig6}b, the mouse IgG with concentrations of 0.625, 1.25, 2.5, 5, 10 ug/mL is subsequently pumped into the flow cell for 50 min after a 10 min circulation of PBS. In Fig. \ref{fig6}c, the model of Langmuir adsorption is applied to fit the experimental results. The absorption constant is calculated as $K_a=6553\ (\mathrm{g/mL})^{-1}$, that is 9829 $\mathrm{M}^{-1}$. It is close to results in other reported literatures \cite{wang2017sensitive,yu2007quantitative,chevrier2004sensitive}. According to the inset figure in Fig. \ref{fig6}b, the standard deviation was calculated to be $5.3\times 10^{-3}$ nm. Based on the three-fold standard deviation theory, the limit of detection (LOD) is calculated to be 15 ng/mL, which is 2 times lower than that of the gold-based SPR biosensor \cite{wang2017sensitive}, which is about 30 ng/mL.

In order to verify the specific binding ability of the sensor, the sensor is tested to detect three different solutions of PBS, human IgG, and mouse IgG. After rabbit anti-mouse IgG is immobilized on the sensor surface, three different solutions are separately circulated into the flow cell. The concentration of human IgG and mouse IgG is 10ug/mL. Fig. \ref{fig6}d illustrates the central wavelength shift of the different solutions after 50 min of injection. It can be seen that there is no obvious central wavelength shift for PBS and human IgG. But for mouse IgG, there is a drastic central wavelength shift. The results have shown that the proposed biosensor has great specificity to detect mouse IgG.

\section{Conclusion}

In this work, an ultrasensitive RI sensor based on the biased weak measurement is proposed to enhance the sensor performance of a total internal reflection system. By introducing an additional phase in the post-selection, the biased weak measurement achieves stronger post-selection and RI sensitivity. Since sensitivity is an important evaluation parameter for the sensor, RI resolution can be further improved with biased weak measurement. In our experiments, our proposed RI sensor with biased weak measurement achieves a RI sensitivity of 13605 nm/RIU. The biased weak measurement strategy improves the RI resolution of the total reflection system to $4\times10^{-7}$ RIU, which is better than those of conventional total internal reflection sensors \cite{chiu2004d} and the standard weak measurement \cite{zhang2016optical}. Based on the three-fold standard deviation theory, the system achieves a detection limit of 15 ng/mL of mouse IgG, which is twice lower than the conventional gold film SPR. Recently, standard weak measurements were applied to the interferometer, SPR \cite{prajapati2021photonic,xu2020measurement,luo2017precision} to improve the performance of the system. This work provides a simple strategy superior to the standard weak measurement, which can be applied to various sensors to enhance their sensing performance.
\section{ORCID}
Chongqi Zhou: 0000-0003-4723-1992
Yanhong Ji:0000-0002-7570-1480

\section*{Acknowledgement}
This research is made possible with the financial support from National Science Foundation of China (NSFC) (81871395, 61875102), Tsinghua University Spring Breeze Fund (2020Z99CFZ023), Oversea cooperation foundation, Graduate School at Shenzhen, Tsinghua University (HW2018007) and Science and Technology Research Program of Shenzhen City (JCYJ20180508152528735, JCYJ20170817111912585).
\bibliography{ref}
\end{document}